\PassOptionsToClass{10pt}{revtex4-1}
\documentclass[aps,prl,showpacs,floatfix,twocolumn,byrevtex,superscriptaddress]{revtex4-1}
\usepackage{amsmath}
\usepackage{amssymb}
\usepackage{amstext}
\usepackage{amsopn}
\usepackage{amsfonts}
\usepackage{amsxtra}
\usepackage[english]{babel}
\usepackage{graphicx}
\usepackage{bm}
\usepackage{multirow}
\usepackage{dcolumn}
\usepackage{color}
\usepackage{hyperref}
\usepackage{todonotes}
\usepackage{verbatim}
\usepackage{soul}
\begin{document}

\title{Observation of Double Weyl Phonons in Parity-Breaking FeSi} 

\author{H. Miao}\email[]{hmiao@bnl.gov}
\affiliation{Condensed Matter Physics and Materials Science Department, Brookhaven National Laboratory, Upton, New York 11973, USA}
\author{T. T. Zhang}
\affiliation{Beijing National Laboratory for Condensed Matter Physics, Institute of Physics, Chinese Academy of Sciences, Beijing 100190, China}
\author{L. Wang}
\affiliation{University of Chinese Academy of Sciences, Beijing 100049, China}
\author{D. Meyers}
\affiliation{Condensed Matter Physics and Materials Science Department, Brookhaven National Laboratory, Upton, New York 11973, USA}
\author{A. H. Said}
\affiliation{Advanced Photon Source, Argonne National Laboratory, Argonne, Illinois 60439, USA}
\author{Y. L. Wang}
\affiliation{Condensed Matter Physics and Materials Science Department, Brookhaven National Laboratory, Upton, New York 11973, USA}
\author{Y. G. Shi}
\affiliation{Beijing National Laboratory for Condensed Matter Physics, Institute of Physics, Chinese Academy of Sciences, Beijing 100190, China}
\author{H. M. Weng}
\author{Z. Fang}
\affiliation{Beijing National Laboratory for Condensed Matter Physics, Institute of Physics, Chinese Academy of Sciences, Beijing 100190, China}
\affiliation{Collaborative Innovation Center of Quantum Matter, Beijing, China}
\author{M. P. M. Dean}\email[]{mdean@bnl.gov}
\affiliation{Condensed Matter Physics and Materials Science Department, Brookhaven National Laboratory, Upton, New York 11973, USA}

\date{\today}

%  72.15.-v  Electronic conduction in metals and alloys
%  74.70.-b  SC: Superconducting materials other than cuprates
%  78.20.-e  Optical properties of bulk materials and thin films
%  78.30.-j  Infrared and Raman spectra
%  74.70.Xa  Pnictides and chalcogenides
%  74.25.Dw  Superconductivity phase diagrams
%  71.10.Hf  Non-Fermi-liquid ground states, electron phase diagrams and phase transitions in model systems

%\pacs{74.70.Xa,74.25.Jb,74.20.Pq}

\date{\today}

\begin{abstract}
Condensed matter systems have now become a fertile ground to discover emerging topological quasi-particles with symmetry protected modes. While many studies have focused on Fermionic excitations, the same conceptual framework can also be applied to bosons yielding new types of topological states. Motivated by the recent theoretical prediction of double-Weyl phonons in transition metal monosilicides  [Phys. Rev. Lett. $\mathbf{120}$, 016401 (2018)], we directly measured the phonon dispersion in parity-breaking FeSi using inelastic x-ray scattering. By comparing the experimental data with theoretical calculations, we make the first observation of double-Weyl points in FeSi, which will be an ideal material to explore emerging Bosonic excitations and its topologically non-trivial properties.

\end{abstract}

\maketitle
In crystalline materials, phonons are the most fundamental Bosonic quasi-particle (QP) that are used to describe the collective motion of the periodically distributed atoms. Inspired by the development of topological band theory on Fermionic QPs \cite{Hasan2010,Qi2011,Armitage2018}, topological phonons have attracted extensive theoretical and experimental attention \cite{Prodan2009,Kane2013,Chen2014,Yang2015,Wang2015,Xiao2015,Nash2015,Susstrunk2015,Mousavi2015,Susstrunk2016,Lu2016,He2016,Fleury2016,Huber2016,Zhang2018}. Previous studies of topological acoustic QPs have focused on artificial macroscopic mechanical systems in the kHz ($\sim10^{-8}$~meV) energy range due to its potential applications in phonoic waveguides \cite{Prodan2009,Kane2013,Chen2014,Yang2015,Wang2015,Xiao2015,Nash2015,Susstrunk2015,Mousavi2015,Susstrunk2016,Lu2016,He2016,Fleury2016,Huber2016}. Very recently, topological phonons have been theoretically  predicted to exist in natural crystalline materials $M$Si ($M=Fe, Co, Mn, Re, Ru$) \cite{Zhang2018}. These topological excitations are the fundamental QPs of the corresponding crystalline lattices and occur in the THz ($\sim10$~meV) energy range, in which phonons often play a dominant role in the thermal and electronic properties of solids. Here we use inelastic x-ray scattering (IXS) to perform the first experimental measurement of the THz topological phonon dispersion in parity-breaking crystalline FeSi. By directly tracking the bulk phonon dispersion near the high symmetry points, we demonstrate the existence of the theoretically predicted double-Weyl points, which have not yet been observed experimentally in related Fermion systems. Our results thus establish transition metal monosilicides as new model systems to explore topological Bosonic excitations and symmetry protected properties such as the topological non-trivial edge modes.

%%%%%%%%%%%%%%%%%%%%%%%%%%%%%%%%%%%%%%%%%%
% Figure 1
%
\begin{figure}[tb]
\includegraphics[width=8 cm]{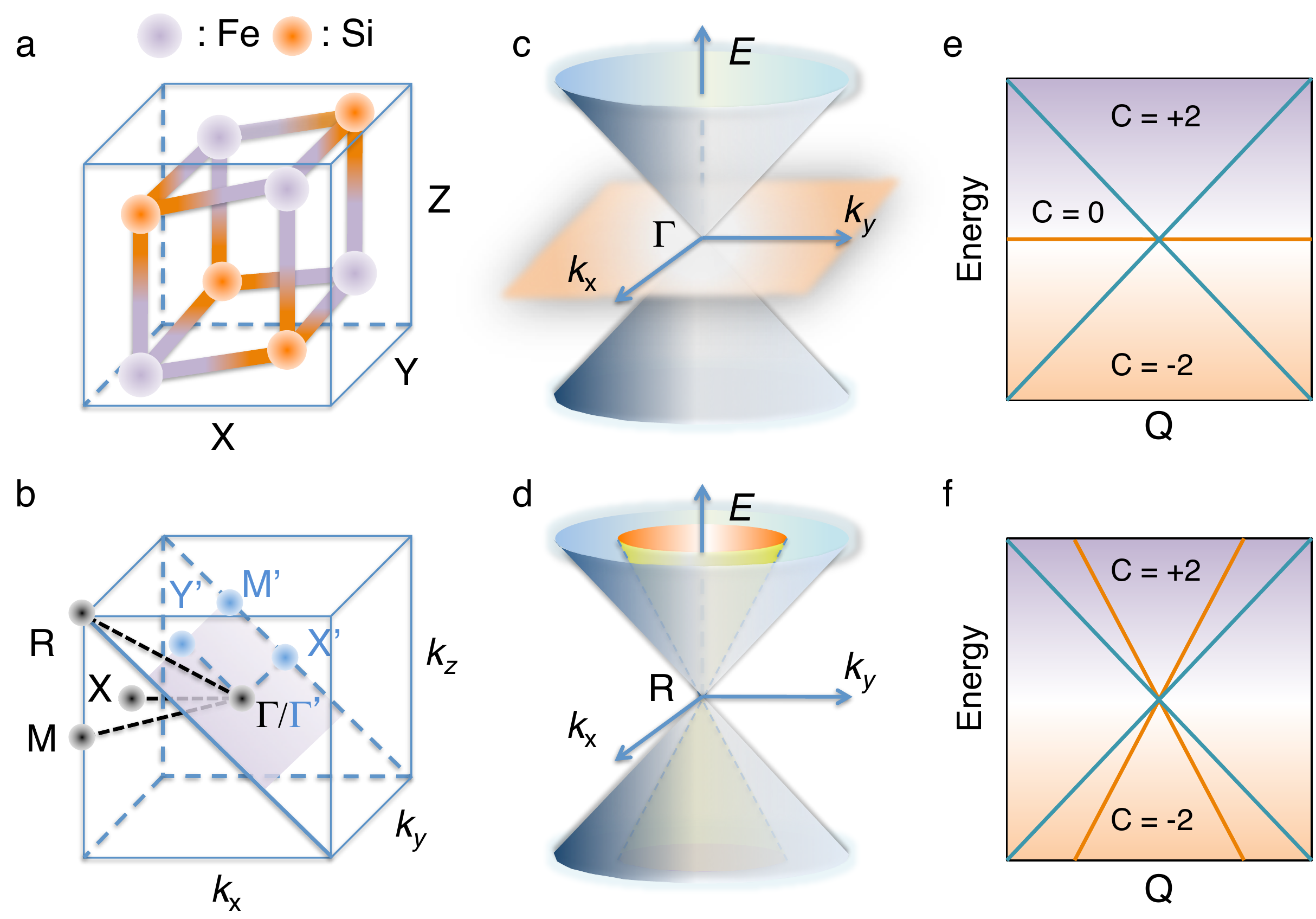}
\caption{Two types of double-Weyl points are predicted in FeSi \cite{Zhang2018}. (a) and (b) show the cubic unit cell and the Brillouin zone (BZ) of FeSi, respectively. The purple $(101)$ plane corresponds to the exposed sample surface in our measurement. The high symmetry points in the 3D BZ and the projected 2D BZ are shown in black and light-blue, respectively. (c) and (d) show the schematic 3D view of the spin-$1$ Weyl point and charge-$2$ Dirac point, respectively. Their corresponding 2D view projections and Chern numbers are shown in (e) and (f), respectively.}
\label{Fig1}
\end{figure}
%%%%%%%%%%%%%%%%%%%%%%%%%%%%%%%%%%%%%%%%%%%

%%%%%%%%%%%%%%%%%%%%%%%%%%%%%%%%%%%%%%%%%%
% Figure 2
%
\begin{figure*}[tb]
\includegraphics[width=13 cm]{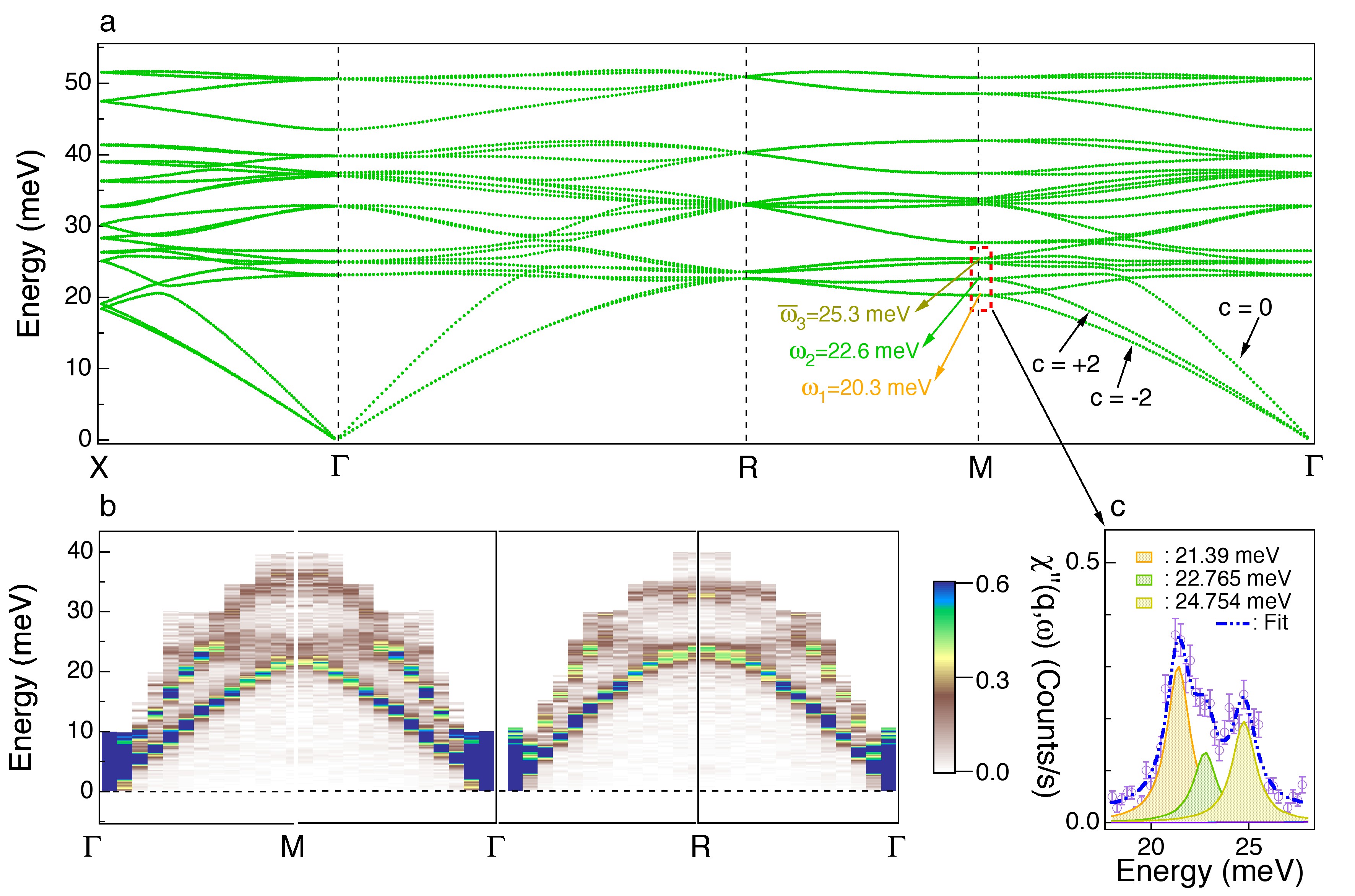}
\caption{Effective spin-1 Weyl acoustic phonons. (a) DFPT calculated phonon dispersion along the $X-\Gamma-R-M-\Gamma$ direction. The Chern numbers for the longitudinal mode at higher energy and the two transverse modes at lower energy are $0$ and $\pm2$, respectively. Due to the parity symmetry ($\mathcal{P}$) breaking in FeSi, the two transverse acoustic modes are split with the largest energy separation located in the red dashed square at the $M$ point. (b) IXS measured phonon dispersion along the $M (4.5, 0.5, 0)-\Gamma (4, 0, 0)-R (4.5, 0.5, 0.5)$ direction. The dispersion along the $\Gamma-M$ and $R-\Gamma$ directions are symmetrized from the raw data with respect to the $M$ and the $R$ point, respectively. (c) High statistics IXS data provide evidence for splitting of the two transverse acoustic phonons at the $M$ point.}
\label{Fig2}
\end{figure*}
%%%%%%%%%%%%%%%%%%%%%%%%%%%%%%%%%%%%%%%%%%%

FeSi has the B20-type structure with noncentrosymmetric space group $P2_{1}3$ (No.~198). As we show in Fig.~\ref{Fig1}a, both Fe and Si atoms are located at the 4(a)-type site in the simple cubic unit cell with position coordinates ($u,u,u$), ($\frac{1}{2}+u,\frac{1}{2}-u,-u$), ($-u,\frac{1}{2}+u,\frac{1}{2}-u$) and ($\frac{1}{2}-u,-u,\frac{1}{2}+u$), where $u_\text{Fe}=0.13652$ and $u_\text{Si}=0.8424$. Therefore, FeSi has three two-fold screw rotations, $C_{2i}'$ ($i=x,y,z$) along each $\langle100\rangle$ axis (e.g.~\{$C_{2x}|(\frac{1}{2},\frac{1}{2},0)$\} along the $x$-axis) and one three fold rotations, $C_{3}$, along the $\langle111\rangle$ axes. In reciprocal space, the particular crystal symmetry of FeSi gives rise to threefold representations at the Brillouin zone (BZ) center (the $\Gamma$ point) and fourfold representations at the BZ corner (the $R$ point as shown in Fig.~\ref{Fig1}b) \cite{Zhang2018}. 

For spin-$1/2$ excitations that are relevant to Fermionic electronic band dispersions, the Weyl-point is usually described by the two-band Hamiltonian, $H_{2}(\textbf{k})=\frac{\hbar}{2}\textbf{k}\cdot\mathbf{\sigma}$. Here $\textbf{k}$ and $\mathbf{\sigma}$ are the QP's momentum and Pauli's matrix, respectively. The corresponding Chern numbers of the two bands are $+1$ and $-1$, respectively. The effective Hamiltonian describing the Weyl-points in the Bosonic three and four band systems can be written as $H_{3}(\textbf{q})=\frac{\hbar}{2}\textbf{q}\cdot\textbf{L}$ and $H_{4}(\textbf{q})=\begin{pmatrix} 
\textbf{q}\cdot\mathbf{\sigma} & 0 \\ 0 & \textbf{q}\cdot\mathbf{\sigma} 
\end{pmatrix}$, respectively. Here $\textbf{q}$ is the reduced momentum transfer in the first Brillouin zone, that is defined as $\textbf{q}=\textbf{Q}-\textbf{G}_{n}$, with $\textbf{Q}$ and $\textbf{G}_{n}$ representing the momentum transfer and reciprocal vector, respectively. $\textbf{L}$ is the spin-$1$ matrix representation of the rotation generator. The Weyl-points and the corresponding Chern numbers that are derived from the effective Hamiltonians $H_{3}(\textbf{q})$ and $H_{4}(\textbf{q})$ are thus dubbed the effective spin-1 Weyl point and the effective charge-$2$ Dirac point respectively, both of which are special cases of double-Weyl points \cite{Zhang2018}.

%%%%%%%%%%%%%%%%%%%%%%%%%%%%%%%%%%%%%%%%%%
% Figure 3
%
\begin{figure*}[tb]
\includegraphics[width=17.2 cm]{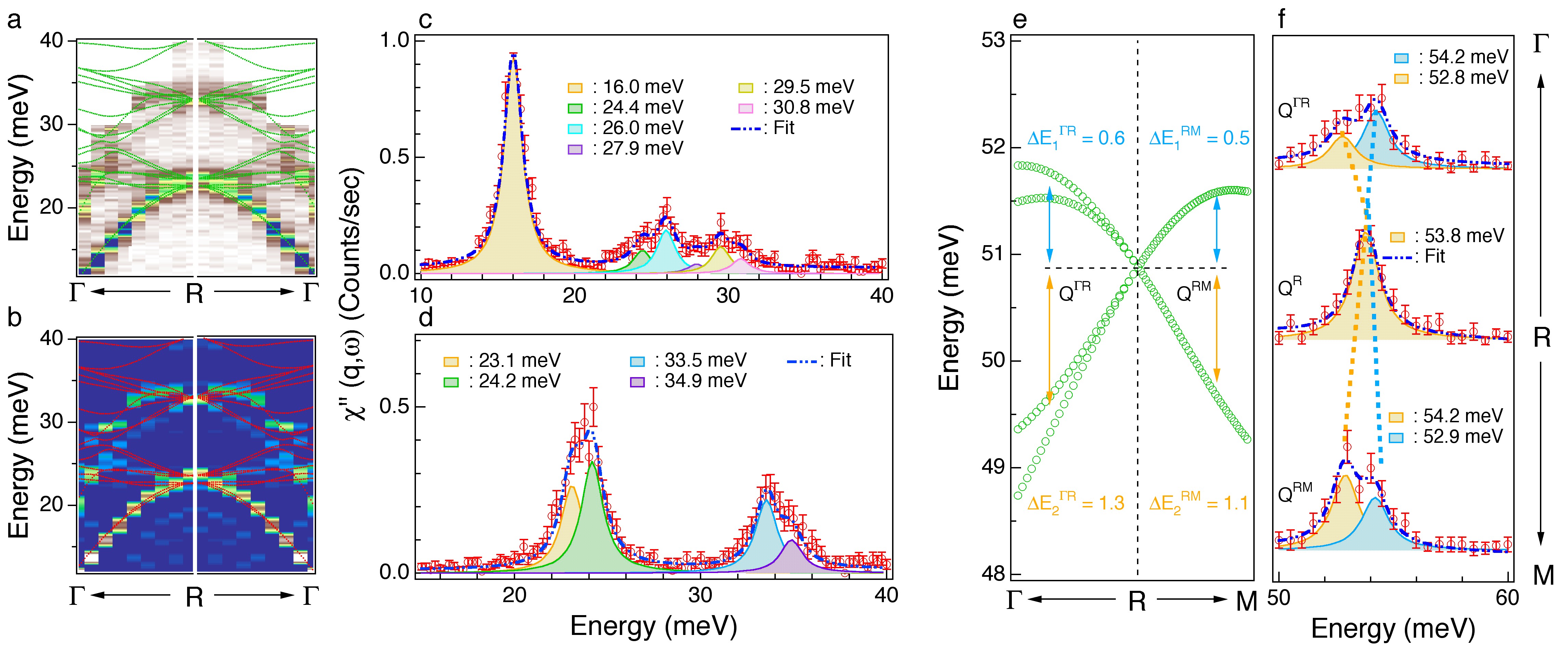}
\caption{Effective charge-2 Dirac points. (a) and (b) Phonon dispersion and its second derivative plot near the $R$ point. Green and red curves are the same re-scaled DFPT calculations. High-statistic IXS spectra at $\mathbf{Q}=(4.26,0.26,0.26)$ and $(4.5, 0.5, 0.5)$ are shown in (c) and (d), respectively. The dashed blue curves are the fitted result. The number of phonon peaks in the fitting is based on the DFPT calculation \cite{Supp}. (e) Re-scaled phonon dispersion near the high-energy effective charge-2 Dirac point. The high-statistic IXS spectra at $\mathbf{Q}^{\Gamma R}$, $\mathbf{Q}^{R}$ and $\mathbf{Q}^{RM}$ are shown in (f). The peak positions extracted from the IXS spectra is slightly higher than the re-scaled DFPT calculation, suggesting different re-scale factors for the low and high-energy phonons.}
\label{Fig3}
\end{figure*}
%%%%%%%%%%%%%%%%%%%%%%%%%%%%%%%%%%%%%%%%%%%

In this letter, we use high-resolution IXS to directly measure the acoustic effective spin-$1$ Weyl point and optical effective charge-$2$ Dirac points in a high quality single crystal FeSi of approximately 2~mm $\times$ 3~mm $\times$ 120~$\mathrm{\mu m}$. The experiments were conducted at beamline 30-ID-C (HERIX \cite{Said2011,Toellner2011}) at the Advanced Photon Source (APS). The highly monochromatic
x-ray beam of incident energy $E_{i}=23.7$~keV ($\lambda = 0.5226$\AA{}) was focused on the sample with a beam cross section of $\sim35\times15$~$\mathrm{\mu m}^{2}$ (horizontal $\times$ vertical). The total energy resolution of the monochromatic x-ray beam and analyzer crystals was $\Delta E\sim 1.5$~meV (full width at half maximum). The measurements were performed in transmission geometry. Typical counting times were in the range of 30 to 120 seconds per point in the energy scans at constant momentum transfer $\textbf{Q}$.

We first focus on the effective spin-1 Weyl acoustic phonons. Figure~\ref{Fig2}a shows the phonon band structure along X-$\Gamma$-R-M-$\Gamma$ direction calculated using VASP \cite{Kresse1996} and density functional perturbation theory (DFPT) \cite{Gonze1997} with fully-relaxed lattice parameters $a=b=c=4.373$ \AA. The generalized gradient approximation was used for the exchange-correlation function \cite{Baroni2001}. We note that phonons in FeSi will be generally softened when increasing temperature or when including spin-phonon interactions \cite{Krannich2015}. Since our calculations were done for the non-magnetic ground state at zero temperature, a re-scaled factor of 0.85 in energy was used to directly compare the IXS data that was performed at room temperature. We note that this scaling is larger than one expects and may reflect an incomplete description of the strong correlations in FeSi \cite{Krannich2015}. Although the three branched acoustic phonon is a generic feature of three dimensional systems, as we shown in Fig.~\ref{Fig2}(a), Chern numbers of the two transverse acoustic modes are well defined only when they are separated from each other, which can only happen if the parity symmetry, $\mathcal{P}$, or the time reversal symmetry, $\mathcal{T}$, is broken \cite{Zhang2018}. Figure~\ref{Fig2}b shows the phonon dispersion along the $M-\Gamma-R$ direction at room temperature. In agreement with the DFPT calculation, the transverse and longitudinal acoustic modes are well separated along both the $\Gamma-M$ and $\Gamma-R$ directions. To prove the two transverse acoustic modes are indeed separated, we performed high-statistic IXS measurement at the $M$ point $(4.5, 0.5, 0)$, where the energy difference is predicted to be the largest in the entire BZ. The data is shown in the energy window between 18 to 28~meV, which corresponds to the red dashed squares shown in Fig.~\ref{Fig2}(a). Three non-degenerate modes are observed in the experimental data that correspond to the well separated transverse acoustic modes and the two nearly degenerate optical modes near 25.3~meV. The IXS spectra were fitted to the standard damped harmonic oscillator functions \cite{Miao2018,Supp}. The results are shown in Fig.~\ref{Fig2}(c) and reveal good consistency with the DFPT calculation.

%Since our measurement is performed at room temperature, the phonon energy is expected to be ``softer'' than the zero-temperature DFPT calculations \cite{Krannich2015}. As shown below, a re-scaling factor of 0.85 happens to overlap the energy with the experimental dispersion, therefore, all bulk and surface phonon calculations shown in this paper are re-scaled in energy by a factor of 0.85. We note that this scaling is larger than one expects and may reflect an incomplete description of the strong correlations in FeSi \cite{Krannich2015}.  
 
%and $\pm$2 for the longitudinal mode and two transverse modes \cite{Zhang2018}. Chern numbers are well defined only when the three acoustic band are separated, which only happens if the parity symmetry, $\mathcal{P}$, or the time reversal symmetry, $\mathcal{T}$, is broken \cite{Zhang2018}. As we have shown in Fig.~\ref{Fig2}a, while the longitudinal mode is well separated from the transverse modes along all high-symmetry directions, the two transverse modes are close in energy with the largest energy separation of 2.2 meV at the M point. 

Having the effective spin-1 Weyl acoustic phonons established, we now turn to the effective charge-2 Dirac points at the $R$ point. The generating elements of the little group at the $R$ point are the rotation symmetry $C_{3}$, the screw rotation symmetry \{$C_{2x}|(\frac{1}{2},\frac{1}{2},0)$\} and the time-reversal symmetry $\mathcal{T}$. The symmetry analysis has proven that the minimal representation is four dimensional with $\mathcal{T}$-protected Chern numbers $\pm2$ \cite{Zhang2018}. Here we first look at the charge-2 Dirac points below 40~meV. Figure~\ref{Fig3}a displays the phonon dispersion near the $R$ point. Its second derivative shown in Fig.~\ref{Fig3}b enables a better visibility of the dispersion that allows direct comparison of the results with the DFPT calculation. The experimentally determined dispersion is in good agreement with the DFPT calculation and resolves both the multiple Charge-2 Dirac points at the $R$ point and the band hybridization along the $\Gamma-R$ direction. To quantitatively determine the band hybridization and the energy of charge-2 Dirac points, we fit the long counting time IXS spectra at $\mathbf{Q}=(4.26,0.26,0.26)$ and the $R$ point (see Figs.~\ref{Fig3}c and d \cite{Supp}). The fitted results are consistent with the DFPT calculation and reveal four charge-2 Dirac points between 20 to 40~meV.

We now move to the high-energy charge-2 Dirac point at the $R$ point. Since the phonon structure factor, $S(q,\omega)$, is proportional to $1/\omega_{q}$, where $\omega_{q}$ is the phonon energy, we restricted our measurements to $\mathbf{Q}^{\Gamma R}$, $\mathbf{Q}^{R}$ and $\mathbf{Q}^{RM}$ shown in Fig.~\ref{Fig3}(f). Based on the DFPT calculation shown in Fig.~\ref{Fig3}(e), the single phonon peak at $\mathbf{Q}^{R}$ is predicted to separate into two and four peaks at $\mathbf{Q}^{RM}$ and $\mathbf{Q}^{\Gamma R}$, respectively. However, due to the small energy difference and the finite equipment resolution, two resolution limited peaks are expected to be observed at both $\mathbf{Q}^{RM}$ and $\mathbf{Q}^{\Gamma R}$. This is indeed what we observed in Fig.~\ref{Fig3}(f): the resolution limited peak at $\mathbf{Q}^{R}$ and $\omega_{R}$=53.8~meV splits into two resolution limited peaks at $\mathbf{Q}^{\Gamma R/RM}$ and $\omega^{\Gamma R/RM}_{up}=54.2/54.2$~meV and $\omega^{\Gamma R/RM}_{down}=52.8/52.9$~meV and, hence, confirm the high-energy charge-2 Dirac point.

%%%%%%%%%%%%%%%%%%%%%%%%%%%%%%%%%%%%%%%%%%
% Figure 4
%
\begin{figure}[tb]
\includegraphics[width=8 cm]{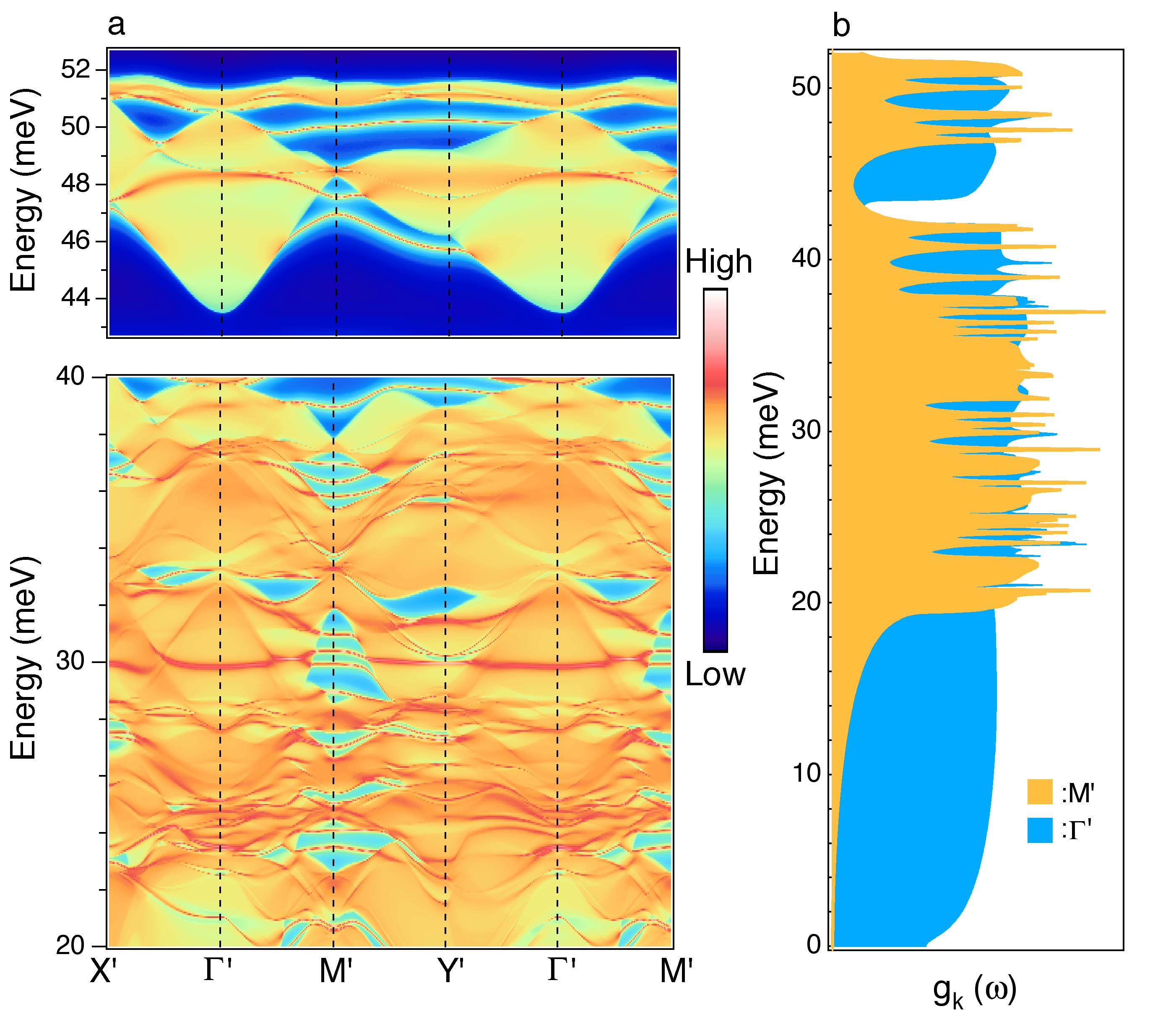}
\caption{Topological surface modes. (a) Momentum resolved phonon dispersion projected on the $(110)$ surface. The definition of the surface high symmetry points is shown in Fig.~\ref{Fig1}(b). (b) Phonon DOS at the surface $M^{\prime}$ and the $\Gamma$' points. The high surface DOS at the $M^{\prime}$ point, making this point as the best place to identify topological edge modes.}
\label{Fig4}
\end{figure}
%%%%%%%%%%%%%%%%%%%%%%%%%%%%%%%%%%%%%%%%%%%

The existence of spin-1 Weyl points at the $\Gamma$ point and the charge-2 Dirac points at the $R$ point will give rise to topologically non-trivial edge modes, which can not be continuously deformed into an even number of non-contractable loops wrapping around the BZ torus. Since x-rays have a penetration depth of a few microns, IXS has minimal sensitivity to the topological phonon modes on the surface. Here, we calculate the (110) surface phonon band structure by constructing the surface Green's function \cite{Sancho1984, Sancho1985, Wu2018, Zhang2018}. Due to the presence of the phonon band gap around 42~meV, the surface states that correspond to the high-energy double-Weyl points are well separated from the bulk band structure through the entire BZ and form a non-contractable Fermi-arc around 50~meV. In contrast, the surface states between 20 and 40~meV are largely overlapping with the bulk band structure making it difficult to be experimentally separated away from the $M^{\prime}$ point. This can be better visualized in Fig.~\ref{Fig4}(b), where the surface phonon density of states (DOS), $g_{M^{\prime}}(\omega)$ and $g_{\Gamma^{\prime}}(\omega)$, are shown at the surface $M^{\prime}$ and the $\Gamma^{\prime}$ point. The calculation at the $M^{\prime}$ point reveals that spikes that correspond to large surface phonon DOS are well separated from the broad continuum, while at the $\Gamma^{\prime}$ point the surface DOS is dominated by the broad continuum. Since the phonon intensity is much larger at low energy, the theoretically predicted high surface phonon DOS between 20 and 40~meV at the surface $M^{\prime}$ point, makes the topological edge modes easier to be identified by surface sensitive probes such as electron energy loss spectroscopy (EELS) and helium scattering.    

In summary, using IXS and theoretical calculations on parity-breaking FeSi, we report the first experimental observation of phonon double-Weyl points. Our results thus establish transition metal monosilicides as new model systems to explore emerging quantum excitations and imply the existence of topological non-trivial edge modes based on the symmetry propagation of the bulk modes.

\begin{acknowledgements}
H.M.\ and M.P.M.D.\ acknowledge A. Alexandradinata, C. Fang, L. Lu and D. Mazzone for insightful discussions. This material is based upon work supported by the U.S. Department of Energy, Office of Basic Energy Sciences, Early Career Award Program under Award No. 1047478. Work at Brookhaven National Laboratory was supported by the U.S. Department of Energy, Office of Science, Office of Basic Energy Sciences, under Contract No. DE-SC00112704. The IXS experiment were performed at 30ID in the Advanced Photon Source, a U.S. Department of Energy (DOE) Office of Science User Facility operated for the DOE Office of Science by Argonne National Laboratory under Contract No. DE-AC02-06CH11357. We acknowledge the supports from the National Key Research and Development Program of China (Grant No. 2016YFA0300600), the National Natural Science Foundation of China (Grant No. 11674369 and No. 11774399), and the Chinese Academy of Sciences (XDB07020100 and QYZDB-SSW-SLH043). H. Miao, T. T. Zhang and L. Wang contributed equally to this work.

\end{acknowledgements}

\bibliography{ref}

\end{document}